\documentclass{ws-ijmpa}
\usepackage[super,compress]{cite}
\usepackage{graphicx}
\begin{document}
\markboth{Sipaz Sharma}{Persistence of charmed hadrons in QGP from lattice QCD}

%
\catchline{}{}{}{}{}
%

\title{Persistence of charmed hadrons in QGP from lattice QCD}

\author{Sipaz Sharma\footnote{For the HotQCD Collaboration.
}
}

\address{Fakult\"at f\"ur Physik, Universit\"at Bielefeld,\\ D-33615 Bielefeld, Germany}



\maketitle


\begin{abstract}
We study the nature of charm degrees of freedom near the chiral crossover based on lattice QCD calculations of the second and fourth-order cumulants of charm fluctuations, and their correlations with net baryon number, electric charge, and strangeness fluctuations. We show that below the chiral crossover temperature, $T_{pc}$, the thermodynamics of charm can be very well understood in terms of charmed hadrons.  We present evidence that above $T_{pc}$,  however, charm quark-like excitations emerge as new degrees of freedom contributing to the partial charm pressure. Nonetheless, up to temperatures as high as 175 MeV, charmed hadron-like excitations significantly contribute to the partial charm pressure. 
\keywords{QCD thermodynamics, charm degrees of freedom, deconfinement}
\end{abstract}

\ccode{PACS: 11.10.Wx, 11.15.Ha, 12.38.Aw, 12.38.Gc, 12.38.Mh, 24.60.Ky, 25.75.Gz, 25.75.Nq}


\section{Introduction}	

At low temperatures and low baryon densities, strongly interacting matter is confined, and the relevant degrees of freedom are hadrons, which are the composite states of the fundamental particles -- quarks and gluons. At low baryon densities, the chiral crossover, ${T_{pc}=156.5\pm1.5}$ MeV \cite{HotQCD:2018pds, Borsanyi:2020fev}, and the creation of Quark-Gluon Plasma (QGP) from hadrons are commonly expected to occur at the same temperature. However, contrary to earlier expectations, recent lattice QCD explorations hint at the survival of some form of hadronic excitations inside the QGP. These excitations may either be thermally modified (broadened) resonances or more complicated quasi-particle excitations. Not much is currently known about the properties of such hadronic excitations. In the open-charm sector, a study by the FASTSUM Collaboration suggests a relatively early modification (at a lower temperature) of the spectral functions in the scalar and axial-vector channels compared to the pseudoscalar and vector channels \cite{Aarts:2022krz}.

The existence of S-wave charmonium (hidden-charm) states within the QGP has been argued based on the relatively small in-medium modifications of screening masses compared to P-wave charmonium states \cite{Bazavov:2014cta}. Another recent study examined the spectral functions of pseudo-scalar light and strange mesons, identifying mesonic bound states at one temperature above the chiral crossover. Evidence for the persistence of pionic excitations in QGP has been found in this analysis \cite{Bala:2023iqu}. The fact that charmed hadrons start melting at $T_{pc}$ has been confirmed over the years by comparison of lattice QCD calculations with Hadron Resonance Gas (HRG) model predictions \cite{BAZAVOV2014210, Sharma:2022ztl}. As the heavier charm quark allows for comparisons of lattice QCD with HRG model calculations within the Boltzmann approximation, the open charm sector offers simplifications. Recent investigations by the HotQCD collaboration based on charm fluctuations and their correlations with other conserved charges have provided supporting evidence for the coexistence of open-charm hadrons and charm quarks inside QGP  \cite{Bazavov:2023xzm,Sharma:2024ucs}. However, the research outcome discussion was mainly centered around the evidence of deconfinement from lattice QCD in terms of the appearance  of charm quark-like excitations above $T_{pc}$. In this article, we will provide additional evidence for the persistence of the charmed hadron-like excitations inside QGP.

\section{Boltzmann Approximation}
\subsection{Charmed Hadrons}
Hadron resonance gas model (HRG) has been successful in describing the particle abundance ratios measured in the heavy-ion experiments. It describes a non-interacting gas of hadron resonances, and therefore can be used to calculate the hadronic pressure below ${T_{pc}}$ \cite{ANDRONIC2019759}. In the Boltzmann approximation, the dimensionless partial pressures from the charmed-meson, ${M_C},$  and the charmed-baryon, ${B_C}$, sectors
	take the following forms \cite{Allton:2005gk}:
	\begin{gather}
		\begin{aligned}
			{M_C(T,\overrightarrow{\mu})}&{=\dfrac{1}{2\pi^2}\sum_{i\in \text{C-mesons}}g_i \bigg(\dfrac{m_i}{T}\bigg)^2K_2(m_i/T)\text{cosh}(Q_i\hat{\mu}_Q+S_i\hat{\mu}_S+C_i\hat{\mu}_C)} \text{ ,}\\
			{B_C(T,\overrightarrow{\mu})}&={\dfrac{1}{2\pi^2}\sum_{i\in \text{C-baryons}}g_i \bigg(\dfrac{m_i}{T}\bigg)^2K_2(m_i/T)\text{cosh}(B_i\hat{\mu}_B+Q_i\hat{\mu}_Q+S_i\hat{\mu}_S+C_i\hat{\mu}_C)} \text{ .}
			\label{eq:McBc}
		\end{aligned}
	\end{gather}
	
	In above equations, at a temperature ${T}$, the summation is over all charmed mesons/baryons (C-mesons/baryons ) with masses given by ${m_i}$; degeneracy factors of the states with equal mass and same quantum numbers are represented by ${g_i}$; to work with a dimensionless notation, chemical potentials corresponding to conserved quantum numbers are normalised by the temperature: ${\hat{\mu}_X = \mu/T}$, ${\forall X \in \{B, Q, S, C\}}$; ${K_2(x)}$ is a modified Bessel function, which for a large argument can be approximated by
	${K_2(x)}\sim\sqrt{\pi/2x}\;e^{-x}\;[1+\mathbb{O}(x^{-1})]$\cite{Allton:2005gk}: consequently, if a charmed state under consideration is much heavier than the relevant temperature scale, such that ${m_i\gg T}$, then the contribution to ${P_C}$ from that particular state will be exponentially suppressed, e.g., the singly-charmed
	${\Lambda}_c^{+}$ baryon has a Particle Data Group (PDG) mass of about $2286$ MeV, whereas the doubly-charmed ${\Xi_{cc}^{++}}$ baryon's mass as tabulated in PDG records is about $3621$ MeV, therefore at ${T_{pc}}$, the contribution to ${B_C}$ from ${\Xi_{cc}^{++}}$ will be suppressed by a factor of $10^{-4}$ in relation to ${\Lambda}_c^{+}$ contribution.
	
	\subsection{Charm Quarks}
	Charm quarks offer an advantage over the light quarks because for temperatures a few times $T_{pc}$, the Boltzmann approximation works for an ideal massive quark-antiquark gas \cite{Allton:2005gk,BAZAVOV2014210}. Therefore, in this approximation, the dimensionless partial charm quark pressure, $Q_C$ is given by, 
	\begin{equation}
		Q_C(T,\overrightarrow{\mu})=\dfrac{3}{\pi^2}\bigg(\dfrac{m_c}{T}\bigg)^2K_2(m_c/T)\text{cosh}\bigg(\dfrac{2}{3}\hat{\mu}_Q+\dfrac{1}{3}\hat{\mu}_B+\hat{\mu}_C\bigg) \text{ ,}
	\end{equation}
	where $m_c$ is the pole mass of the charm quark, and the degeneracy factor is 6.
	\begin{figure}[h]
		\centering
	\includegraphics[width=0.8\textwidth, height=7.3 cm]{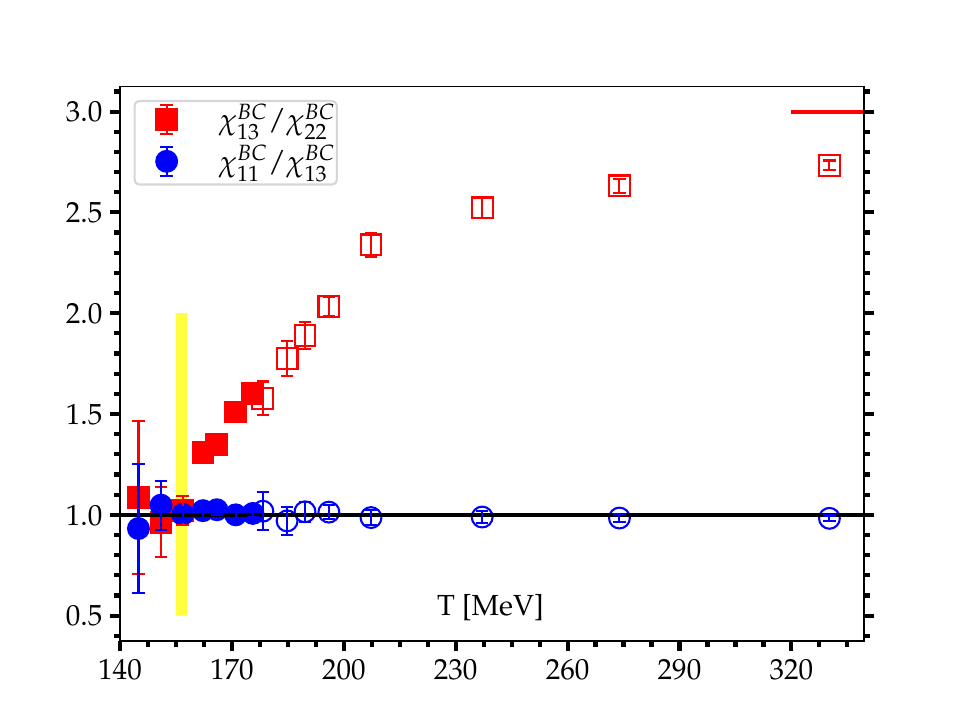}
	\caption{The ratios of different baryon-charm fluctuations as functions of temperature. The open symbols
		represent the results from Ref. \cite{BAZAVOV2014210}.
		The red solid line is the ideal charm quark gas limit of the ratio $\chi_{13}^{BC}/\chi_{22}^{BC}$. The yellow band represents $T_{pc}$ with its uncertainty. This figure is taken from \cite{Sharma:2024ucs}.}
	\label{fig:HRG}
\end{figure}
	\section{Generalized Susceptibilities of the Conserved Charges}
	\label{sec: susc}
	To project out the relevant degrees of freedom in the charm sector, one calculates the generalized susceptibilities, ${\chi^{BQSC}_{klmn}}$, of the conserved charges. This involves taking appropriate derivatives of the total pressure ${P}$, which contains contributions from the total charm pressure, $P_C(T,\vec{\mu})=\chi_2^C=\chi_n^C$, for $n$ even. These derivatives are taken with respect to the chemical potentials of the quantum number combinations one is interested in:
	\begin{equation} 
		{\chi^{BQSC}_{klmn}=\dfrac{\partial^{(k+l+m+n)}\;\;[P\;(\hat{\mu}_B,\hat{\mu}_Q,\hat{\mu}_S,\hat{\mu}_C)\;/T^4]}{\partial\hat{\mu}^{k}_B\;\;\partial\hat{\mu}^{l}_Q\;\;\partial\hat{\mu}^{m}_S\;\;\partial\hat{\mu}^{n}_C}}\bigg|_{\overrightarrow{\mu}=0}\text{.}
		\label{eq:chi}
	\end{equation}
	
	To make R.H.S dimensionless, $P$ is normalized by $T^4$. If ${n \neq 0}$, ${P}$ can be replaced by ${P_C}$, since the derivative w.r.t. ${\hat{\mu}_C}$ will always project onto the open-charm sector. Note that ${\chi^{BQSC}_{klmn}}$ will be non-zero only for ${(k+l+m+n) \in \text{even}}$. In the following, if the subscript corresponding to a conserved charge is zero in the L.H.S. of Eq. \ref{eq:chi}, then both the corresponding superscript as well the zero subscript will be suppressed. Also, the terms cumulants, fluctuations and generalized susceptibilities will be used interchangebly throughout the text.
\begin{remark}
	We show results for lattices with a fixed temporal extent, i.e., $N_{\tau}=8$, and with a fixed aspect ratio of 4. Our conclusions are based on observables constructed using suitable ratios of various generalized susceptibilities. The lattice cutoff effects cancel up to a large extent in ratios; therefore, we expect our conclusions to hold in the continuum limit. To corroborate this claim, a few results for  $N_{\tau}=12$ lattices are also presented. For further details, please refer to \cite{BAZAVOV2014210, Sharma:2022ztl}.
\end{remark}

\begin{remark}
	Certain ratios of charm fluctuations calculated in the framework of lattice QCD can receive enhanced contributions due the existence of not-yet-discovered open-charm states. Therefore, to achieve an agreement with the lattice data, one needs to take into account these extra states to make HRG predictions.  In the following, all the HRG predictions, in addition to states tabulated in the Particle Data Group (PDG) records, take into account states predicted via Quark-Model calculations \cite{Ebert:2009ua, Ebert:2011kk, Chen:2022asf}. These HRG calculations are denoted by QM-HRG.
	\end{remark}

\section{Breakdown of the HRG description}	

	In the validity range of HRG, all $BC$ correlations project onto the partial pressure contribution from the charmed baryonic sector, $B_C$. Therefore, the ratios of $BC$ correlations with different numbers of $\hat{\mu}_B$ derivatives, such as $\chi^{BC}_{13}/\chi^{BC}_{22}$ shown in Fig.~\ref{fig:HRG}, give a clear indication of hadron melting by deviating from unity. Slightly above $T_{pc}$, $\chi^{BC}_{13}/\chi^{BC}_{22}$ becomes greater than $1$, signaling the appearance of charm degrees of freedom carrying fractional $B$. For $T>300 \text{MeV}$, $\chi^{BC}_{13}/\chi^{BC}_{22}$ approaches its non-interacting quark-gas limit, which is $3$ -- shown with the solid-red line. Due to the dominance of the singly-charmed sector, $\chi^{BC}_{11}/\chi^{BC}_{13}$ shown in Fig.~\ref{fig:HRG} stays unity for the entire temperature range.
	
	\section{Charmed degrees of freedom above $T_{pc}$}
	
	To understand the nature of charm degrees of freedom above $T_{pc}$, we extend the simple hadron gas model allowing the presence of partial charm quark pressure based on Ref.~\cite{Mukherjee:2015mxc}:
	\begin{align}
		P_C(T,\vec{\mu})=M_C(T,\vec{\mu})+B_C(T,\vec{\mu})+Q_C(T,\vec{\mu}) \, .
		\label{eq:Pmodel}
	\end{align}
	For details please see our recent work \cite{Bazavov:2023xzm}. By considering only two quantum numbers: $B$ and $C$, the partial pressures of quark, baryon and meson-like excitations for $\mu=0$ can be expressed in terms of the generalized susceptibilities as follows,
	
	\begin{align}
		P_{q}&=9(\chi^{BC}_{13}-\chi^{BC}_{22})/2\; , 
		\label{eq:partial-quasi} 	\\
		P_{B}&=(3\chi^{BC}_{22}-\chi^{BC}_{13})/2\; , 
		\label{eq:partial-quasiB}\\
		P_{M}&=\chi^{C}_{4}+3\chi^{BC}_{22}-4\chi^{BC}_{13} \; .
	\end{align}
	
	\subsection{Appearance of Quark-Like Excitations near $T_{pc}$}

\begin{figure}[h]
		\includegraphics[width=0.49\textwidth]{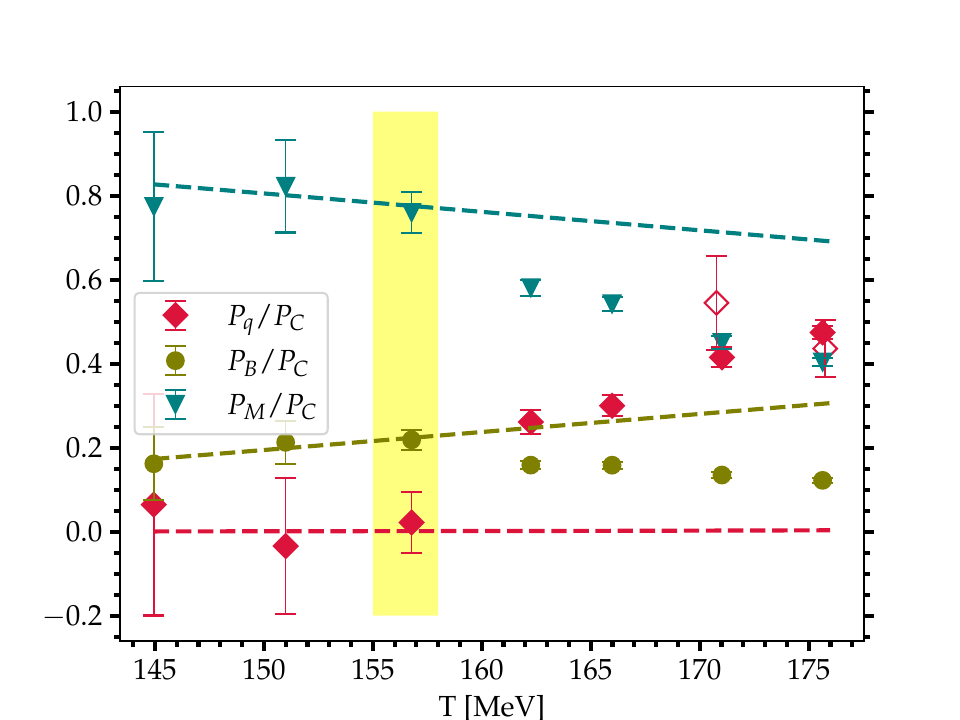}
		\includegraphics[width=0.49\textwidth]{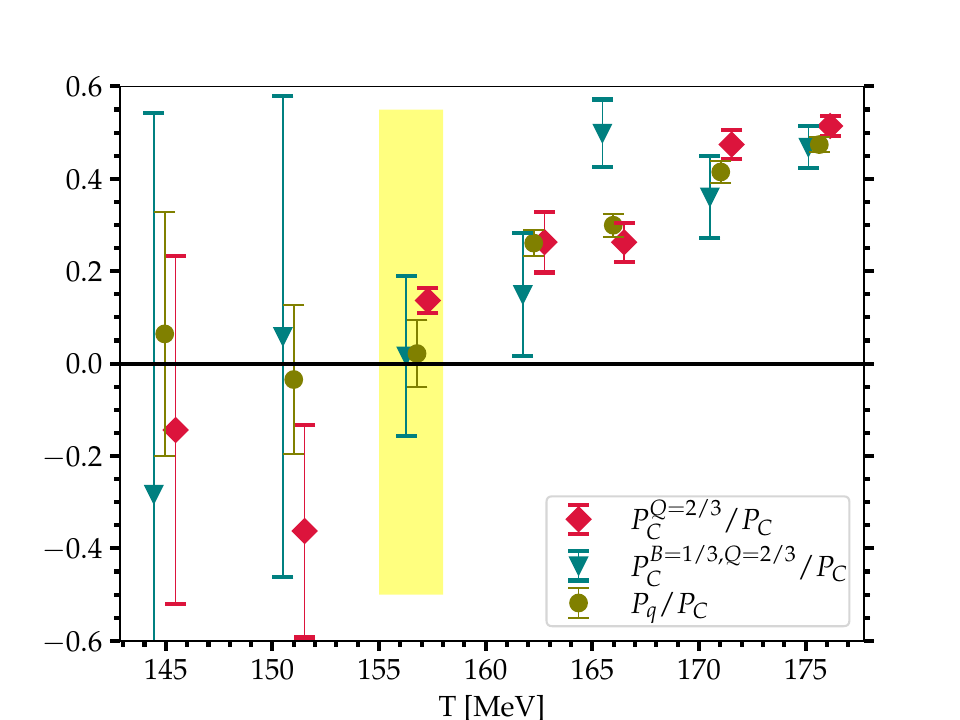}
		\caption{ Partial pressures of charmed mesons, charmed baryons and charm quarks normalized by $P_C$ as functions of temperature. The dashed lines show corresponding results obtained from the QM-HRG model. The open symbols show the results for $N_{\tau}=12$ lattices [left]. The partial pressures 
			of quasi-particles carrying
			(i) baryon number $|B|=1/3$ ($P_q$), (ii) electric charge
			$|Q|=2/3$ ($P_C^{Q=2/3}$), and
			(iii) ($|B|=1/3,\ |Q|=2/3$)  ($P_C^{B=1/3,Q=2/3}$) normalized by $P_C$.
			Note that the T-coordinates of case (ii) and case (iii) are shifted by $\pm 0.51$ MeV respectively [right]. The yellow bands represent $T_{pc}$ with its uncertainty. These figures are taken from \cite{Sharma:2024ucs}.}
		\label{fig:quarks}
	\end{figure}

	Upon breakdown of the HRG description at $T_{pc}$ in Fig. \ref{fig:quarks}~[left], the fractional contribution of both the charmed mesonic and the charmed baryonic states to the total charm pressure starts decreasing in comparison to their respective QM-HRG expectations, whereas the fractional contribution of the charmed states with $|B|=1/3$ becomes non-zero slightly above $T_{pc}$, and continues to increase as a function of temperature. To rule out the role of lattice cutoff effects in making $P_q/P_C$ non-zero, we also show $N_\tau=12$ results for the highest two temperatures using unfilled-red markers in Fig. \ref{fig:quarks}~[left]. The $N_\tau=12$  results clearly agree with the $N_\tau=8$ results within errors, which further supports the presence of charm quark-like excitations in the Quark-Gluon Plasma (QGP).
	To strengthen our claim, in addition to the partial quark pressure construction using $B$ and $C$ quantum numbers in Eq. \ref{eq:partial-quasi}, we independently construct partial pressure of quark-like excitations in two other ways. Firstly, by considering two quantum numbers: $Q$ and $C$. Secondly, by considering three quantum numbers: $B$, $Q$ and $C$. In addition to $P_q/P_C$, both these partial pressures normalised by $P_C$ are shown in Fig. \ref{fig:quarks}~[right], and they take the following forms:
	\begin{align}
		P_{C}^{Q=2/3}&=\frac{1}{8}\big[54\chi^{QC}_{13}-81\chi^{QC}_{22}+27\chi^{QC}_{31}\big]\; ,\\
		P_{C}^{B=1/3,Q=2/3}&=\frac{27}{4}\big[\chi^{BQC}_{112}-\chi^{BQC}_{211}]\; .
	\end{align}
	Above partial pressures are constructed by assuming that only states carrying $|B|=0,1\text{ and }1/3$ contribute to $P_C$. This assumption is justified because our lattice data within errors satisfies the condition, $\chi_{13}^{BC} - 4\chi_{22}^{BC} + 3\chi_{31}^{BC} = 0$, see Ref.~\cite{Bazavov:2023xzm}. This implies four possibilities in the $QC$ sector: $|Q| = 0, 1, 2 \text{ and }2/3$, and three possibilities in the $BQC$ sector:  i) $|B|=1, |Q|=1$;  ii) $|B|=1, |Q|=2$;  iii) $|B|=1/3, |Q|=2/3$. Fig. \ref{fig:quarks}~[right] shows a remarkable agreement between three independent partial pressure constructions of quark-like excitations providing support to the quasi-particle model in Eq. \ref{eq:Pmodel}. Notice that $P_{C}^{Q=2/3}$ is sensitive to contributions from $|Q|\neq 0, 1 \text{ and } 2$ charm sectors. The fact that it agrees with $P_q$ and $P_{C}^{B=1/3,Q=2/3}$ implies that these three quantities project onto the same quasi-particle sector.

		\begin{figure}[h]
			\centering
		\includegraphics[width=0.8\textwidth, height=7.3cm]{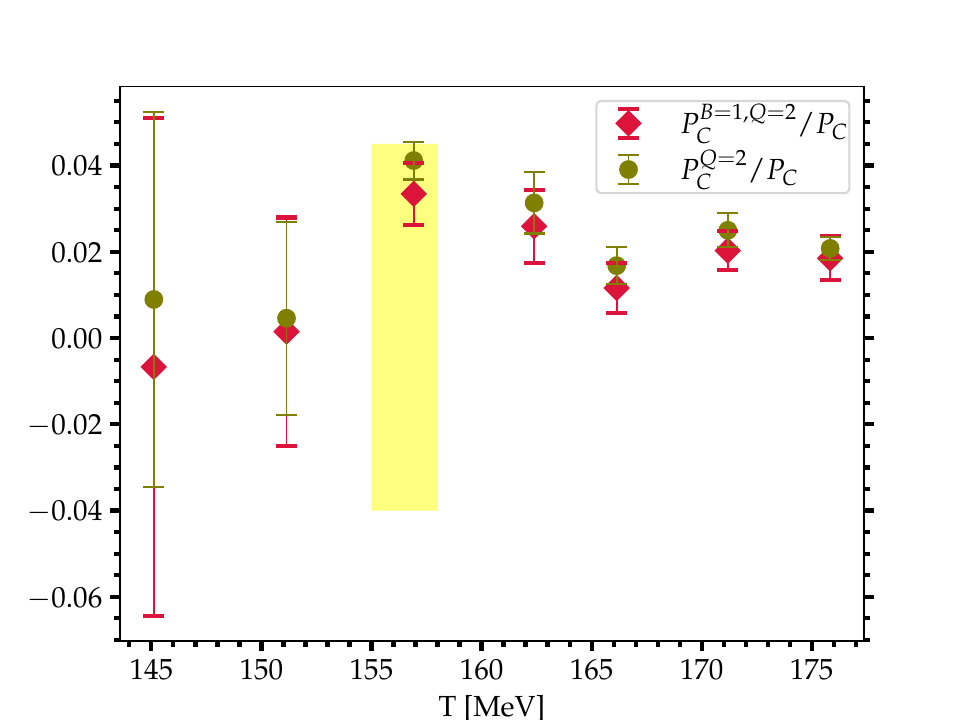}
		\caption{The partial pressures 
			of quasi-particles carrying
			 (i) electric charge
			$|Q|=2$ ($P_C^{Q=2}$), and
			(ii) ($|B|=1,|Q|=2$)  ($P_C^{B=1,Q=2}$) normalized by $P_C$. The yellow band represents $T_{pc}$ with its uncertainty.}
		\label{fig:B1Q2}
	\end{figure}
		\subsection{Persistence of charmed hadrons above $T_{pc}$}
It is evident from Fig~\ref{fig:quarks}~[left] that charmed hadrons remain the dominant contributors to the total charm pressure until $T=175.84 $ MeV. It is worth pointing out that this is the highest temperature explored in this work and does not provide any information on physical thresholds. At this temperature, both quark and hadron degrees of freedom contribute equally to $P_C$, and the charm quark becomes the dominant degree of freedom roughly above $1.12~ T_{pc}$. 

Below, $T_{pc}$, the  $|Q|=2$ charmed sector is solely composed of the charmed baryons. Therefore, if the proposed model in Eq. \ref{eq:Pmodel} holds, then even above $T_{pc}$, only charmed baryons should contribute to the partial pressure from the  $|Q|=2$ charmed sector. It is again possible to independently project onto  the  $|Q|=2$ charmed sector and the $|Q|=2$ charmed baryonic sector using $QC$ and $BQC$ correlations respectively:
\begin{align}
		P_{C}^{Q=2}&=\frac{1}{8}\big[2\chi^{QC}_{13}-5\chi^{QC}_{22}+3\chi^{QC}_{31}\big]\\
		P_{C}^{B=1,Q=2}&=\frac{1}{4}\big[-\chi^{BQC}_{211}+2\chi^{BQC}_{121}-\chi^{BQC}_{112}\big]
	\end{align}
Fig.~\ref{fig:B1Q2} shows a clear agreement between the  above two partial pressure constructions, thereby corroborating the persistence of charmed hadrons above $T_{pc}$. 

\begin{figure}[h]
	\centering
	\includegraphics[width=0.8\textwidth, height=7.8cm]{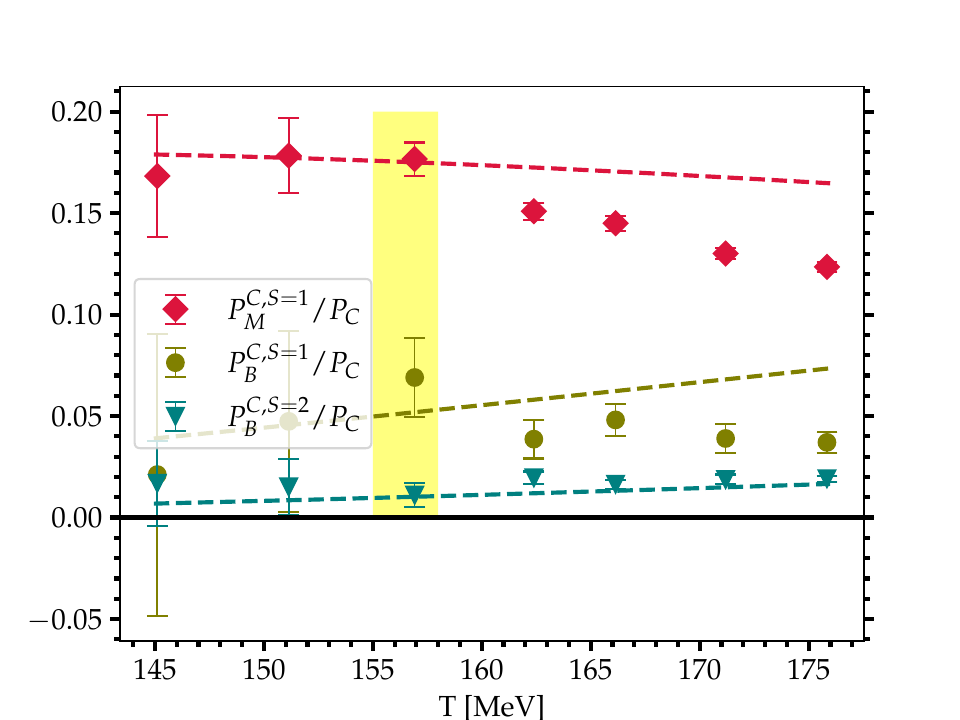}
	\caption{The partial pressures 
		contributions of (i) $|S|=1$ charmed mesonic sector ($P_M^{C,S=1}$),
		(ii) $|S|=1$ charmed baryonic sector  ($P_B^{C,S=1}$), and (iii) $|S|=2$ charmed baryonic sector  ($P_B^{C,S=2}$) normalized by $P_C$. The yellow band represents $T_{pc}$ with its uncertainty. The dashed lines show corresponding results obtained from the QM-HRG model. }
	\label{fig:SC}
\end{figure}

It is important to emphasize that the three out of four constraints satisfied by the proposed quasi-particle model come from the strange-charm sector (see Ref. \cite{Bazavov:2023xzm}). These constraints validate the assumption that only states carrying $|B|=0,1 \text{ and } 1/3$ contribute to $P_C$ above $T_{pc}$. One might think of only using $|B|=1/3$ states to describe the charm thermodynamics above $T_{pc}$. However, such possibility is refuted by the non-zero $SC$ and $BSC$ correlations above $T_{pc}$. Based on the proposed model in Eq. \ref{eq:Pmodel}, the partial pressures from $|S|=1$ charmed mesonic sector, $|S|=1$ charmed baryonic sector, and $|S|=2$ charmed baryonic sector are shown in Fig.~\ref{fig:SC} and take the following forms:

\begin{align}
	P_{M}^{C,S=1}&=\chi_{13}^{SC}-\chi_{112}^{BSC}\\
	P_{B}^{C,S=1}&=\chi_{13}^{SC}-\chi_{22}^{SC}-3\chi_{112}^{BSC}\\
	P_{B}^{C,S=2}&=(2\chi_{112}^{BSC}+\chi_{22}^{SC}-\chi_{13}^{SC})/2
	\end{align}
Therefore, based on Fig.~\ref{fig:SC}, non-zero $SC$ and $BSC$ correlations support coexistence of charmed hadrons and quarks inside QGP.

\section{Conclusions}

We show that the HRG description of the charmed degrees of freedom breaks down at $T_{pc}$. Our results give evidence of deconfinement in terms of the appearance of charm quark-like excitations at $T_{pc}$. We show that the relevant charm degrees of freedom inside  QGP fall into three categories: meson-like, baryon-like and quark-like, and  the charmed hadrons are the dominant degrees of freedom for $T<1.12~ T_{pc}$. Moreover, similar to ${T<T_{pc}}$ regime, our data suggests that for ${T>T_{pc}}$, the ${|Q|=2}$ charmed sector is solely composed of baryon-like states. The non-zero strange-charm correlations provide further support for the persistence of charmed hadrons above $T_{pc}$. Our results approach the ideal charm quark gas limit for $240 \text{ MeV}<T\leq340 \text{ MeV}$. 

\section*{Acknowledgements}
This work was supported by The Deutsche Forschungsgemeinschaft (DFG, German Research Foundation) - Project number 315477589-TRR 211,
”Strong interaction matter under extreme conditions”.
The authors gratefully acknowledge the
computing time and support provided to them on the high-performance computer Noctua 2 at the NHR Center
PC2 under the project name: hpc-prf-cfpd. These are funded by the Federal Ministry of Education
and Research and the state governments participating on the basis of the resolutions of the GWK
for the national high-performance computing at universities (www.nhr-verein.de/unsere-partner).
Numerical calculations have also been performed on the
GPU-cluster at Bielefeld University, Germany. We thank the Bielefeld HPC.NRW team for their support. All the HRG calculations were performed using the AnalysisToolbox code developed by the HotQCD Collaboration \cite{Altenkort:2023xxi}.

\bibliographystyle{unsrt} 
\bibliography{refs}

\end{document}